\documentclass[fp,twocolumn]{jpsj3}
\usepackage{txfonts}
\usepackage{color}
\usepackage{ulem}

\title{Gradual Charge Order Melting in Bi$_{0.5}$Ca$_{0.5}$MnO$_3$ Induced by Ultrahigh Magnetic Fields}

\author{Yuto Ishii$^1$\thanks{yutoishii@issp.u-tokyo.ac.jp}, Akihiko Ikeda$^2$, Masashi Tokunaga$^1$, Koichi Kindo$^1$, Akira Matsuo$^1$, and Yasuhiro H. Matsuda$^1$}
\inst{$^1$Institute for Solid State Physics, University of Tokyo, Kashiwa, Chiba 277-8581, Japan \\
$^2$Departiment of Engineering Science, University of Electro-Communications (UEC), Chofu, Tokyo, 182-8585, Japan} 

\abst{We have investigated the magnetization process of Bi$_{0.5}$Ca$_{0.5}$MnO$_3$ under ultrahigh magnetic fields.
The metamagnetic transition due to the magnetic field-induced charge order melting has been observed.
The metamagnetic transition becomes broad and smeared in the field ascending process at low temperatures.
However, in the field descending process, a clear metamagnetic transition is observed even at low temperatures.
Gradual charge order melting by magnetic field, and its reformation are expected to occur in ascending and descending magnetic fields, respectively.
Quenched spatial inhomogeneity of the charge distribution and/or quantum criticality can account for the observed asymmetric broadening of the field induced charge order melting.}

\begin{document}
\maketitle

\section{Introduction}

Perovskite-type manganites exhibit abundant exotic physical phenomena such as colossal magnetoresistance (CMR), charge/orbital order, ferromagnetic (FM) metallic behavior due to the double exchange mechanism, and multiferroic properties owing to the coupling between the spin, charge, and orbital degrees of freedom \cite{RN102}.
In perovskite-type manganites, Mn ions are located at the center of O$_6$ octahedra, and thus, the five degenerated 3$d$ orbital energy levels split into three degenerated $t_{\rm 2g}$ orbitals and two degenerated $e_{\rm g}$ orbitals.
In Mn$^{3+}$ ($t_{\rm 2g}^3$ $e_{\rm g}^1$) or Mn$^{4+}$ ($t_{\rm 2g}^3$) states, the three $t_{\rm 2g}$ electrons are weakly hybridized with O 2$p$ states, and consequently they possess the localized total spin $S$ = 3/2.
However, the $e_{\rm g}$ electrons can be itinerant due to the strong hybridization with the O 2$p$ orbitals.
The spins of the localized $t_{\rm 2g}$ electron and the conduction $e_{\rm g}$ electron are always aligned with each other due to the strong on-site ferromagnetic coupling (Hund's coupling).
There are several important interactions such as Jahn-Teller coupling, on-site and inter-site Coulomb repulsion, superexchange interaction, and double exchange interaction.
Thus, the balance of these interactions determines the electronic state.
It is known that when a ratio of the number of Mn$^{3+}$ ions and that of Mn$^{4+}$ ions is nearly 1, the charge order (CO) state becomes most stable \cite{RN102}.
The important property is that the CO state is significantly affected by a magnetic field.
The transfer integral of $e_{\rm g}$ electrons between Mn ions increases by a magnetic field because the spins of the $t_{\rm 2g}$ localized electrons align.
As a result, the bandwidth of the $e_{\rm g}$ electron increases and the CO can be melted.

$Ln_{1-x}AE_{x}$MnO$_3$ ($Ln$ : rare earth element, $AE$ : Ca or Sr) series (rare earth perovskite-type manganites) have been systematically investigated \cite{RN98,RN123,RN94,RN124,RN96,RN97,RN125,RN93,RN126,RN127,RN128,RN129}.
With a large ionic radius of the $A$-site ions, for example, ($Ln$, $AE$) = (La, Sr), the tilt of MnO$_6$ octahedra is small, which results in an enhancement of the transfer integral and the bandwidth becomes large.
Therefore, the FM metallic ground state realizes when the $x$ is between roughly 0.15 and 0.5 \cite{RN98,RN123}.
By decreasing the average ionic radius of the $A$-site ions, the tilt becomes large; thus, localized CO/OO states become stable due to a reduction of the transfer integral.
In almost all of the CO phases, the magnetic ground state is antiferromagnetic (AFM) ordered state.
In fact, $Ln$ = (Pr, Nd, Sm) systems show the CO-AFM ordered ground state at low temperatures \cite{RN94,RN124,RN125,RN93,RN126,RN127,RN128,RN129}.
Because of the high stability of the CO state, the critical magnetic field required for the CO melting increases with decreasing the ionic radius of the $A$-site ion \cite{RN96,RN97}.

In perovskite-type manganites, a spatial inhomogeneity of charge distribution is the one of the attractive subjects and also the important factor for the electronic ground state and phase transitions. Actually, some manganites show the coexistence of the CO insulating state and FM metallic state \cite{Mori1998, Katsufuji1999, Moritomo1999}. In addition, the phase separation and its percolation phenomenon are believed to be the origin of the CMR \cite{{Uehara1999}}.

When the $Ln$ site is occupied by the Bi$^{3+}$ ion, the physical properties are different from those of the rare earth manganites \cite{Bokov1967,RN99,RN100,RN77}.
The 6$s^2$ lone pair electrons of Bi$^{3+}$ tend to hybridize with the 2$p$ orbital of the oxygen ion.
In BiMnO$_3$, multiferroic behavior is regarded to be induced by this hybridization \cite{RN83,RN101,RN90}.
By substitution of Ca$^{2+}$ with Bi$^{3+}$, Bi$_{0.5}$Ca$_{0.5}$MnO$_3$ undergoes the CO phase transition at the transition temperature $T_{\rm CO}$ = 325 K \cite{Bokov1967,RN99,RN100,RN77}.
The $T_{\rm CO}$ is higher than that in most of the rare earth systems $Ln_{1-x}AE_{x}$MnO$_3$ with $x$ = 0.5.
The CO phase transition occurs along with the structural transition. It was investigated in detail on Bi$_{0.6}$Ca$_{0.4}$MnO$_3$ by the single-crystal X-ray diffraction \cite{RN81}. A unit cell expansion due to the CO phase transition takes place with the symmetry lowering from $Pnma$ to $Pnm2_1$ \cite{RN81}.
The CO pattern in Bi$_{0.5}$Ca$_{0.5}$MnO$_3$ is the checkerboard-type, which was confirmed by transmission electron microscopy \cite{RN105}.
In addition, the AFM phase transition at $T_{\rm N}$ = 133 K was also confirmed in Bi$_{0.5}$Ca$_{0.5}$MnO$_3$ \cite{Bokov1967,RN77}.
Regarding the high magnetic field study, the magnetization measurement was performed by A. Kirste $et$ $al$. \cite{RN87} and S. Takeyama $et$ $al$. \cite{RN131} up to ultrahigh magnetic fields exceeding 100 T.
They found the metamagnetic transition at the high magnetic field.
These previous studies suggest that the Bi$^{3+}$ ion affects the electronic and magnetic states, as well as the CO phase stability against the magnetic field.
A systematic investigation of the high-field magnetic properties of Bi$_{0.5}$Ca$_{0.5}$MnO$_3$ to uncover the electronic properties would be of great interest.

\begin{figure}
\centering
\includegraphics[width=0.95\columnwidth]{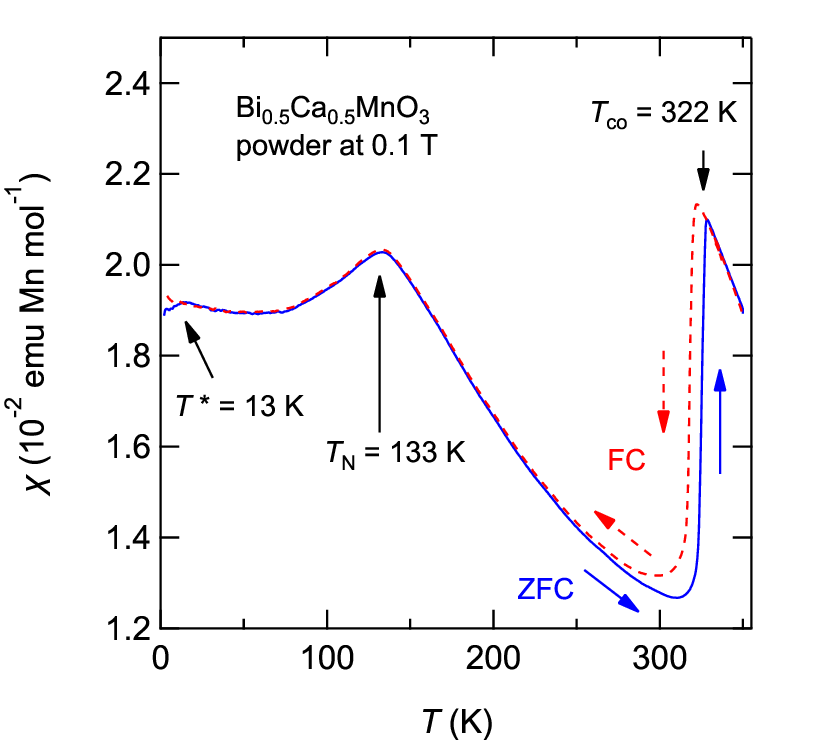}
\caption{\label{Chi-T} (Color online) Magnetic susceptibility of Bi$_{0.5}$Ca$_{0.5}$MnO$_3$ under a magnetic field of 0.1 T as a function of temperature. The blue-solid and red-broken lines represent the zero-field-cooling and field-cooling processes, respectively.}
\end{figure}

In this paper, we study the magnetization of Bi$_{0.5}$Ca$_{0.5}$MnO$_3$ in ultrahigh magnetic fields of up to 140 T over a wide temperature range from 11 to 293 K.
The metamagnetic transition due to the magnetic field-induced CO melting occurs at $B_{\rm c}^{\rm up}$ = 54 T and $B_{\rm c}^{\rm down}$ = 34 T at 293 K, where $B_{\rm c}^{\rm up}$ and $B_{\rm c}^{\rm down}$ are the critical fields in the field ascending and descending processes, respectively.
The critical field varies with the temperature change, and the hysteresis increases with decreasing temperature.
The phase transition becomes significantly broad in the field ascending process below around $T_{\rm N}$.
Meanwhile, even below $T_{\rm N}$, the transition is sharp in field descending process at $B_{\rm c}^{\rm down}$.
This asymmetric change of first-order phase transition is a new finding of this work.
The results can be explained by the magnetic field-induced gradual CO melting and its reformation when the magnetic field becomes weak.
The quenched spatial inhomogeneity of the charge distribution and/or quantum criticality can be a cause of the asymmetric field-induced CO transition.
Finally, we constructed the $B$-$T$ phase diagram of Bi$_{0.5}$Ca$_{0.5}$MnO$_3$ including the unusual critical feature.

\section{Experiments}
Single crystalline samples of Bi$_{0.5}$Ca$_{0.5}$MnO$_3$ were prepared by the Bi$_2$O$_3$ flux method.
Sample characterization was performed by the Rietveld analysis on powder X-ray diffraction data and scanning electron microscopy \cite{supple}.
The temperature dependence of the magnetic susceptibility of Bi$_{0.5}$Ca$_{0.5}$MnO$_3$ was measured using a SQUID magnetometer (Quantum Design MPMS).
The sample was powdered by crushing and grinding the single crystals.
High-field and ultrahigh-field magnetization measurements were also performed using the powder sample.

\begin{figure}
\centering
\includegraphics[width=0.95\columnwidth]{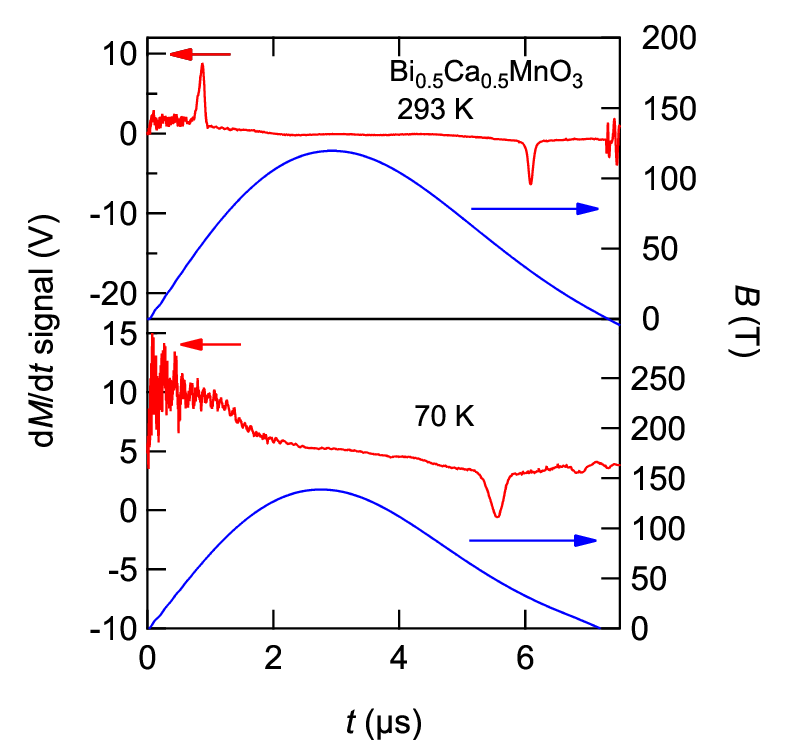}
\caption{\label{data_S3} (Color online) Red lines show the induction voltages proportional to the d$M$/d$t$. Blue lines represent the magnetic field as a function of time. The Upper and bottom panels show the data at $T$ = 293 K and 70 K, respectively. Maximum fields are $B_{\rm max}$ = 120 T and 140 T for shots at $T$ = 293 K and 70 K, respectively.}
\end{figure}

Pulsed high magnetic fields of up to 40 T were generated using a nondestructive pulse magnet at the Institute for Solid State Physics (ISSP), the University of Tokyo.
Pulsed ultrahigh magnetic fields of up to 140 T  were generated by a horizontal-type single-turn coil (HSTC), which is a destructive pulse magnet, at the ISSP \cite{RN107}.
The pulse duration time of the magnetic field is about 7 $\mu$s in HSTC.
Helium flow-type cryostats made of fiber-reinforced plastics were employed for the experiments to avoid generating an induction current with a high sweep rate of the magnetic field \cite{RN108,RN109}.
The sample temperature was monitored by a chlomel-constantan thermocouple.
The magnetization was measured using a pickup coil consisting of two small coils (1 mm diameter, 1.2 mm length for each).
The two coils have opposite polarities and are connected in series.
The sample is set into one of the two coils.
A voltage proportional to the time derivative of magnetization $M$ (d$M$/d$t$) is induced when the sample is magnetized, where $t$ is the time.
An induction voltage due to the time derivative of the magnetic field d$B$/d$t$ is almost canceled out by the two coils.
The residual d$B$/d$t$ induction voltage was removed by subtraction of the induction voltage of the pickup coil without the sample.
The detailed experimental setup for the magnetization measurement using the HSTC method is described in elsewhere \cite{RN108,RN109}.
The absolute value of $M$ was calibrated using the data obtained from the nondestructive pulsed magnetic field experiments and MPMS.
The time evolution of the magnetic field generated by the HSTC was measured using a calibrated pickup coil (0.7 mm diameter, 5 turns) placed near the sample.

\section{Results}
The temperature dependence of the magnetic susceptibility at 0.1 T is shown in Fig. \ref{Chi-T}.
The susceptibility shows a steep drop at the charge order temperature $T_{\rm CO}$ = 322 K and a peak at the N\'{e}el temperature $T_{\rm N}$ = 133 K, which is consistent with the previous studies \cite{Bokov1967,RN77}. The susceptibilities of zero-field-cooling (ZFC) and field-cooling (FC) deviate from each other below $T$ = 13 K.
We define this temperature as another characteristic temperature $T^*$.
Figure \ref{data_S3} shows the time derivative of the magnetization (d$M$/d$t$) signals (red line) and magnetic fields $B$ (blue line) of ultrahigh field magnetization measurements as a function of time at $T$ = 293 K and 70 K.
The maximum magnetic fields are $B_{\rm max}$ = 120 T and 140 T for the experiments at $T$ = 293 K and 70 K, respectively.
At $T$ = 293 K, sharp peaks with positive and negative signs appeared in the field ascending and descending processes, respectively.
In contrast, d$M$/d$t$ at 70 K shows only one peak in the field descending process.

\begin{figure}
\centering
\includegraphics[width=0.95\columnwidth]{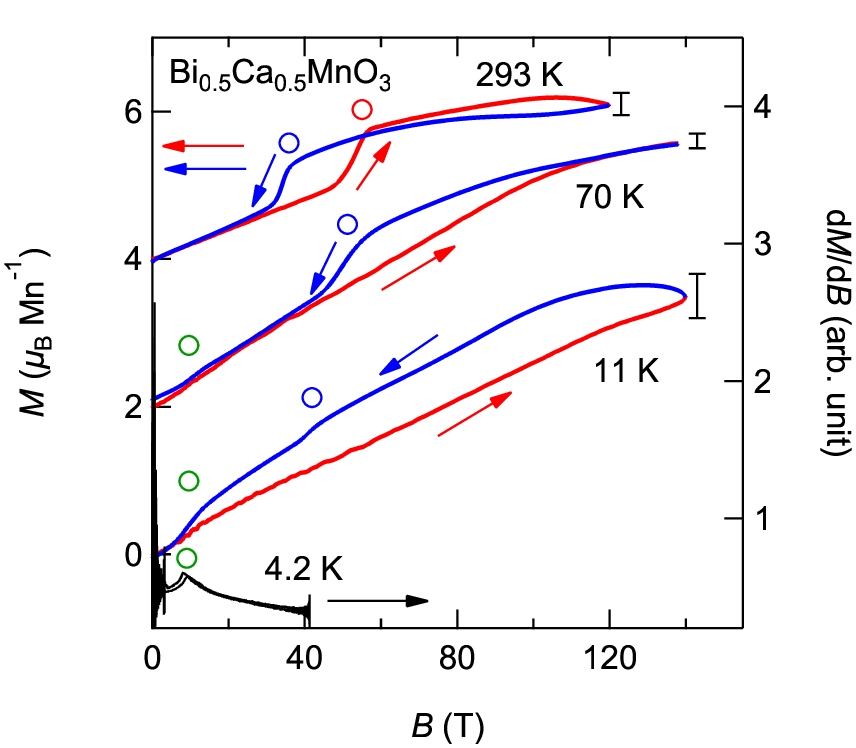}
\caption{\label{MHs_S3} (Color online) Magnetization curves at 293 K (up to 120 T), 70 K (140 T), and 11 K (140 T). Red and blue lines represent the field ascending and descending processes, respectively. Black line is the d$M$/d$B$  at 4.2 K up to 40 T measured by the nondestructive method. Critical fields are marked by red and blue open circles in the field ascending ($B_{\rm c}^{\rm up}$) and descending processes ($B_{\rm c}^{\rm down}$), respectively. Green open circles mark the critical field of the metamagnetic transition which appears around 10 T below $T_{\rm N}$. The estimated error bar is shown for each measurement.}
\end{figure}

\begin{figure}
\centering
\includegraphics[width=0.95\columnwidth]{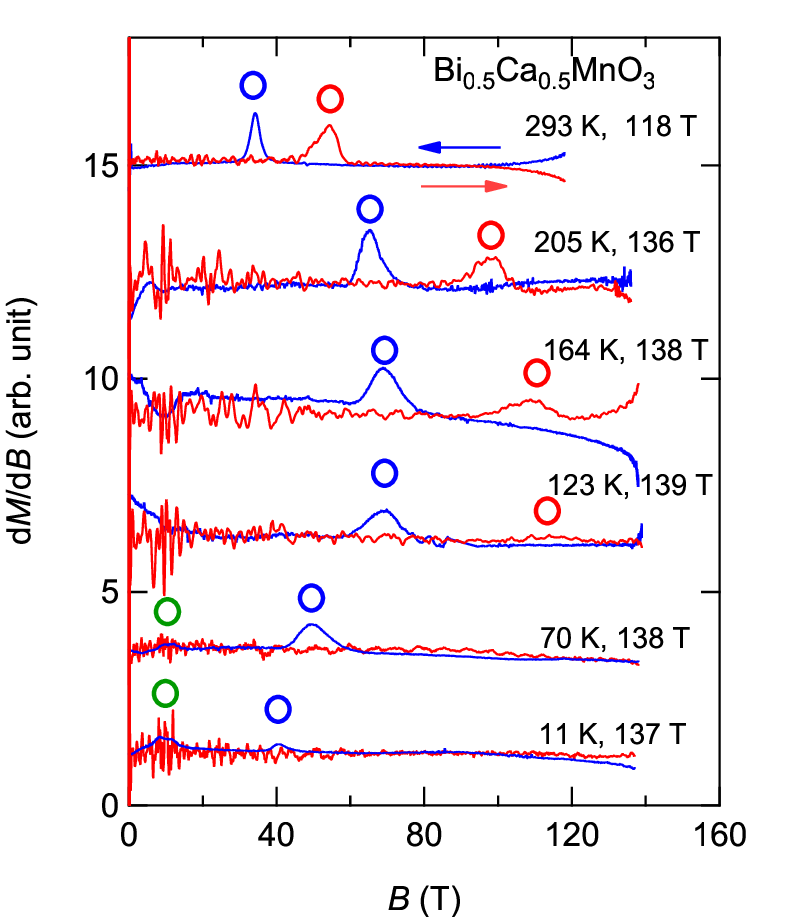}
\caption{\label{dMdB_S3_all} (Color online) Field differentials of the magnetization d$M$/d$B$ of BCMO at several temperatures with each $B_{\rm max}$. Red and blue lines represent the field ascending and descending processes, respectively. Peaks corresponding to the phase transition in the field ascending and descending processes are marked by red and blue open circles, respectively. Small metamagnetic transition at low temperatures is marked by the green open circles.}
\end{figure}

\begin{figure*}[h]
\centering
\includegraphics[width=1.95\columnwidth]{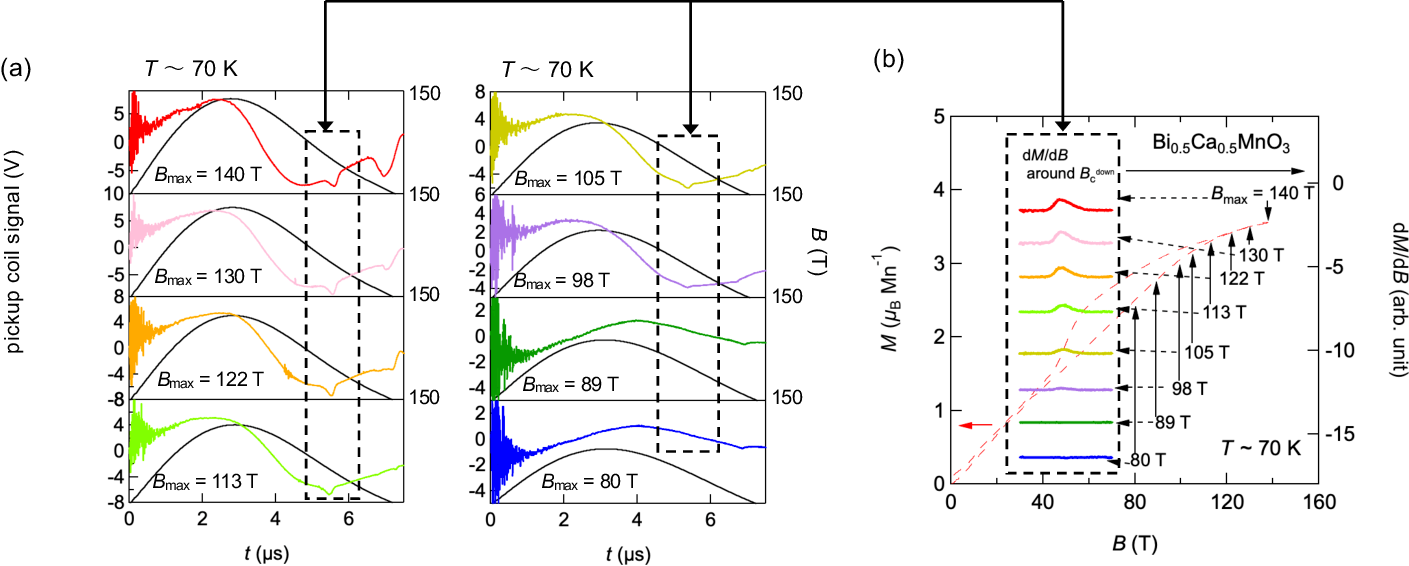}
\caption{\label{BmaxRaw} (Color online) (a) Raw pickup coil signals (color lines) and magnetic fields (black lines) as a function of time. Broken-line rectangles are the field regions around the d$M$/d$B$ peak in the field descending process. (b) Magnetization at 70 K up to 140 T (red broken line) and the d$M$/d$B$ around $B_{\rm c}^{\rm down}$ subtracted by the other contributions at several $B_{\rm max}$ (other color solid lines).}
\end{figure*}

Magnetization curves obtained by numerical integration of d$M$/d$t$ at $T$ = 293 K, 70 K, and 11 K are shown in Fig. \ref{MHs_S3} (red and blue lines).
The d$M$/d$B$ data at $T$ = 4.2 K up to 40 T measured by the nondestructive pulse magnet is also plotted in the same graph (black line).
The red and blue lines are the magnetization curves in the field ascending and field descending processes, respectively.
The magnetization at $T$ = 293 K shows a metamagnetic transition at the critical field $B_{\rm c}^{\rm up}$ = 54 T in the field ascending process and that at $B_{\rm c}^{\rm down}$ = 35 T in the field descending process.
The critical fields reported \cite{RN131} are ($B_{\rm c}^{\rm up}$ = 58 T and $B_{\rm c}^{\rm down}$ = 36 T at $T$ = 290 K) are consistent with those observed in the present study.

At 70 K, no sudden jump was detected in the field ascending process, however, a distinct steep decrease appeared in the field descending process at $B_{\rm c}^{\rm down}$ = 50 T.
The steep decrease at $B_{\rm c}^{\rm down}$ = 41 T is small at 11 K.
This unusual phenomenon is more clearly seen in the temperature variation of the d$M$/d$B$ curve as shown in Fig. \ref{dMdB_S3_all}.

The d$M$/d$B$ peak at $B_{\rm c}^{\rm up}$ becomes broad with decreasing temperature and seems to nearly disappear at lower than 123 K.
Meanwhile, the d$M$/d$B$ peak is clearly observed at $B_{\rm c}^{\rm down}$ when $T >$ 11 K.
The $B_{\rm c}^{\rm down}$ shifts toward the low-field side with decreasing temperature.
At 11 K, the peak at $B_{\rm c}^{\rm down}$ remains but its intensity is small.

In addition, another small metamagnetic transition around 10 T appears below $T_{\rm N}$.
This transition also appeared in nondestructive experiments as shown in Fig. \ref{MHs_S3} (black line).
This is probably due to the spin flop transition coming from the weak anisotropy in the AFM ordered state.

Here, it could be noted that the metamagnetic transitions at $B_{\rm c}^{\rm up}$ = 72 T, and $B_{\rm c}^{\rm down}$ = 40 T were reported at a low temperature in the previous work on the same material \cite{RN87}.
This result is not consistent with the observation in the present study because $B_{\rm c}^{\rm up}$ is much higher and the d$M$/d$B$ peak becomes unclear at low temperatures in the present study. However, the reason is not very clear. And further discussion is difficult because no data on temperature dependence is shown in ref. \cite{RN87}.

\section {Discussion}

Below the $T_{\rm CO}$, a magnetic field is expected to suppress the CO state owing to an enhancement of the electron transfer probability through the alignment of the orientation of the spins of Mn ions \cite{RN102}.
This kind of metamagnetic transition is usually understood as the field-induced CO melting in perovskite-type manganites \cite{RN102}.
The field-induced CO melting accompanies the structural phase transition from the structure below $T_{\rm CO}$ to that above $T_{\rm CO}$, which is confirmed by the X-ray diffraction experiment under the pulsed high magnetic field in Pr$_{0.6}$Ca$_{0.4}$MnO$_3$ \cite{YHM2004}.
In Bi$_{0.5}$Ca$_{0.5}$MnO$_3$, this structural transition was also confirmed by the X-ray diffraction experiment under the high magnetic field of up to 77 T at room temperature (around 300 K) \cite{AIkeda2022}. Therefore, the metamagnetic transition in Bi$_{0.5}$Ca$_{0.5}$MnO$_3$ at room temperature observed in our work and S. Takeyama $et\ al$'s work \cite{RN131} can be understood as the field-induced CO melting.

When the temperature is decreased from 293 K to 123 K, the critical field shifts toward the high-field side, and the hysteresis width gradually increases.
This behavior is qualitatively similar to that of rare earth perovskite-type manganites \cite{RN97}.
It is found that the highest critical field of Bi$_{0.5}$Ca$_{0.5}$MnO$_3$ is higher than 100 T near $T_{\rm N}$ (= 133 K).
This means that the CO in Bi$_{0.5}$Ca$_{0.5}$MnO$_3$ is more stable than that in rare earth perovskite-type manganites of similar Mn$^{3+}$/Mn$^{4+}$ ratio \cite{RN123,RN91,RN96}.
The d$M$/d$B$ peak at $B_{\rm c}^{\rm up}$ is gradually smeared by decreasing the temperature and this smearing becomes remarkable below around $T_{\rm N}$. 
Furthermore, the peak at $B_{\rm c}^{\rm up}$ disappears below 70 K, although the peak at $B_{\rm c}^{\rm down}$ is clear (Fig. \ref{dMdB_S3_all}).

Judging from the behavior of the d$M$/d$B$ peak in the field descending process, it can be said that the high field and low field phases are the CO-melting and the CO phases, respectively, at all the measurement temperatures.
Therefore, the transition from the low-field CO phase to the high-field CO-melting phase must occur in the field ascending process even at low temperatures.
Here, we need to consider how we can define the $B_{\rm c}^{\rm up}$ without using the d$M$/d$B$ peak.
Our idea is to find the $B_{\rm c}^{\rm up}$ by checking whether the d$M$/d$B$ peak in the field descending process appears or not; i.e., if a magnetic field exceeds $B_{\rm c}^{\rm up}$, the d$M$/d$B$ peak in the field descending process should appear, while if the field is lower than $B_{\rm c}^{\rm up}$, there should be no peak.
Therefore, we investigated the $B_{\rm max}$ dependence of the magnetization.

Figure \ref{BmaxRaw} (a) shows the raw signals of pickup coil including the both of d$M$/d$t$ and background (color lines) and magnetic fields as a function of time (black lines) with several $B_{\rm max}$ at $T \sim$ 70 K.
The negative peak at approximately 5.6 $\mu$s corresponds to the phase transition at $B_{\rm c}^{\rm down}$.
The rectangle of broken lines represents the regions around the $B_{\rm c}^{\rm down}$.
The d$M$/d$B$ curves around $B_{\rm c}^{\rm down}$ are shown inside of the broken-line rectangles in Fig. \ref{BmaxRaw} (b), along with the magnetization curve up to 140 T at $T$ = 70 K denoted by a red-broken line.
The background signals were subtracted from the d$M$/d$B$ data.
The d$M$/d$B$ peak appears when $B_{\rm max}$ is higher than 98 T, which suggests the $B_{\rm c}^{\rm up}$ is between 89 T and 98 T.
The peak height increases with increasing $B_{\rm max}$, whereas the position of $B_{\rm c}^{\rm down}$ is constant.

\begin{figure}
\centering
\includegraphics[width=0.7\columnwidth]{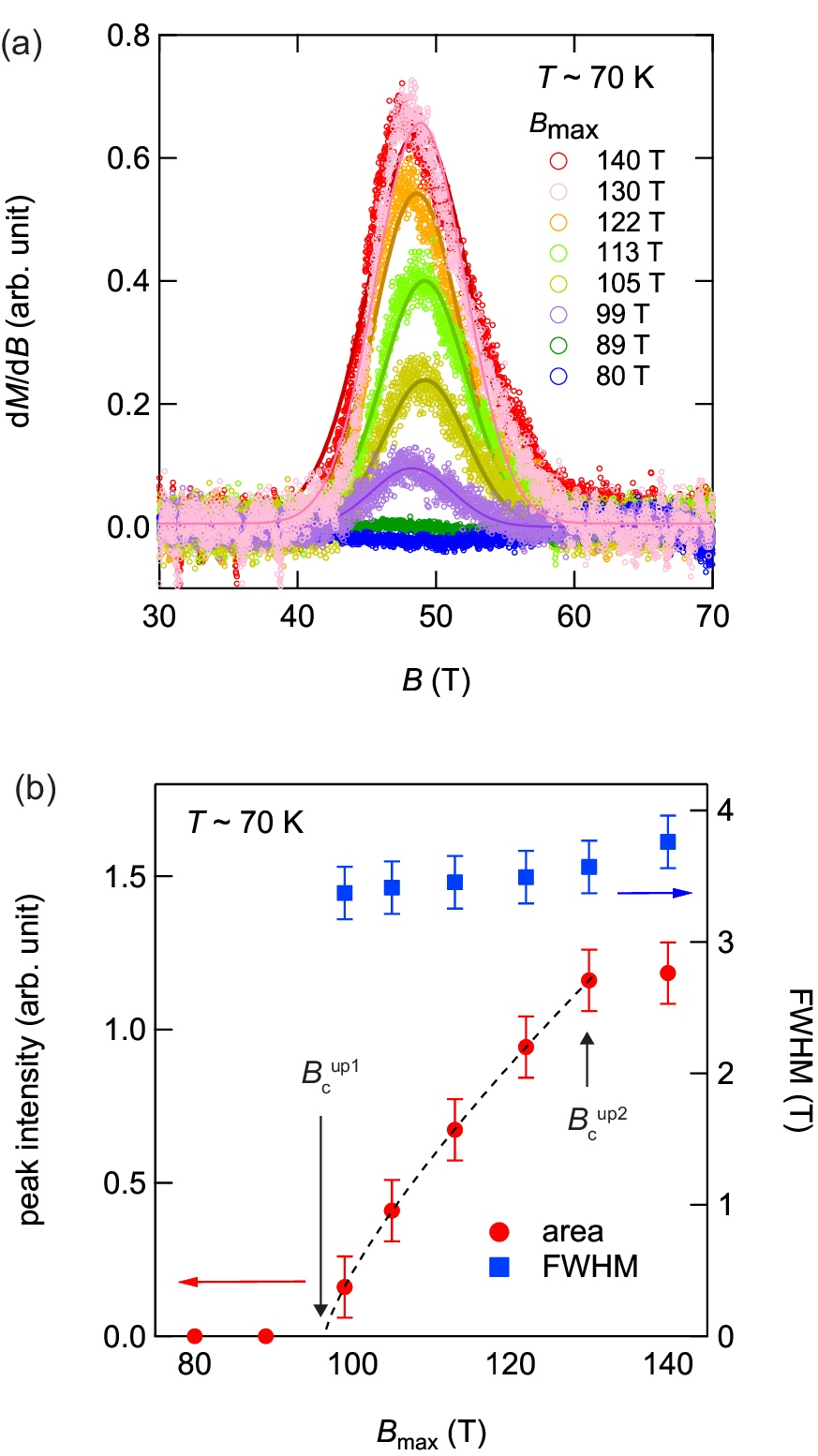}
\caption{\label{PeakFit} (Color online) (a) d$M$/d$B$ around the $B_{\rm c}^{\rm down}$ at $T$ = 70 K with several $B_{\rm max}$ shown with the same baseline. Open circles and solid lines represent the experimental data and Gaussian fitting results, respectively. (b) $B_{\rm max}$ dependence of the intensity and FWHM of the d$M$/d$B$ peak around the $B_{\rm c}^{\rm down}$ in the field descending process.}
\end{figure}

The d$M$/d$B$ peaks shown in Fig. \ref{BmaxRaw} (b) are drawn in Fig. \ref{PeakFit} (a) with the same baseline.
We tried to fit the peaks using a Gaussian function to quantitatively analyze the $B_{\rm max}$ dependence of the peak intensity.
The fitting results are shown in Fig. \ref{PeakFit} (a) as solid lines.
From these, the $B_{\rm max}$ dependence of the d$M$/d$B$ peak intensity (evaluated by the product of the height and width of the peak) and the full width at half maximum (FWHM) were obtained, as shown in Fig. \ref{PeakFit} (b).
The peak intensity increases with $B_{\rm max}$ and saturates at 130 T.
However, the FWHM is almost constant for all $B_{\max}$.
The magnetic field where the peak intensity becomes finite was estimated to be $B$ = 96 T by extrapolating the $B_{\max}$ dependence of the intensity (black broken line).
We define it as the critical field $B_{\rm c}^{\rm up1}$ and the field where the peak intensity saturates as the other critical field $B_{\rm c}^{\rm up2}$.

The metamagnetic transition in the field descending process at $B_{\rm c}^{\rm down}$ is indicative of CO reformation.
Therefore, the fact that the peak at $B_{\rm c}^{\rm down}$ appears only when $B_{\rm max}$ is larger than $B_{\rm c}^{\rm up1}$ = 96 T suggests that the CO starts melting from the $B_{\rm c}^{\rm up1}$.
The continuous increase in the peak intensity at $B_{\rm c}^{\rm down}$ indicates that the volume fraction of the CO melting region increases gradually with increasing $B_{\rm max}$.
This can be understood as that the CO is gradually melted by the magnetic field between $B_{\rm c}^{\rm up1}$ = 96 and $B_{\rm c}^{\rm up2}$ = 130 T (at 70 K).
We can define $B_{\rm c}^{\rm up1}$ and $B_{\rm c}^{\rm up2}$ as the critical fields where the CO starts and completes melting, respectively.
The constant FWHM and $B_{\rm c}^{\rm down}$ suggest that the CO reforms rapidly at the same magnetic field $B_{\rm c}^{\rm down}$.
Based on the results shown in Fig. \ref{dMdB_S3_all}, this behavior is likely to appear below $T_{\rm N}$ (= 133 K), suggesting the development of AFM order can be an important factor for this unusual behavior.
Regarding the small peak at $B_{\rm c}^{\rm down}$ at 11 K (Fig. \ref{dMdB_S3_all}), it is expected that the $B_{\rm c}^{\rm up2}$ increases with decreasing temperature, and 140 T is not high enough to fully induce the CO-melting phase at this temperature.

Here, we discuss the possible origin of the broadening of the phase transition.
The broadening of the metamagnetic transition suggests that the spatial inhomogeneity or the slow dynamics. In addition, the broadening is asymmetric with respect to the way of changing magnetic field (ascending or descending). Therefore, we will discuss including this asymmetry.

Generally, other perovskite-type rare earth manganites don't show the large asymmetric broadening as observed in this work. However, strictly, metamagnetic transitions of some perovskite-type manganites are asymmetric \cite{RN97}. In addition, almost all of high-field magnetization curves at low temperatures reported in M. Respaud $et\ al$ \cite{RN97} shows that the transition in the field descending process is sharper than that in the field ascending process. This tendency is consistent with this work.

According to the previous studies, the spatial inhomogeneity of the charge distribution can occur in manganites \cite{Machida2002}. With decreasing temperature, the spatial inhomogeneity of the charge distribution can be frozen. If it happens, the spatial inhomogeneity of the critical field for the CO melting becomes prominent resulting in the broad field-induced CO melting. In the CO melting region, the spacial inhomogeneity will become less because the electron becomes itinerant. Therefore, reflecting the better homogeneity, the metamagnetic transition in the field descending process can occur sharply.

The magnetic susceptibility shows the deviation between the ZFC and FC processes below $T^{*}$ (Fig. \ref{Chi-T}). 
This observation may indicate a spin-glass like behavior, and thus the spatial inhomogeneity of the magnetic interaction which is supposed to be strongly coupled to the spacial inhomogeneity of the charge distribution.

Another possible explanation is the one based on the slow dynamics.
A critical slowing down near a quantum critical point (QCP) may play an important role.
The QCP of AFM order can be induced by the suppression of the orders by the magnetic field.
The Weiss temperature was estimated as $\theta_{\rm W} \sim$ - 10$^2$ K in the paramagnetic CO phase, which is the same energy scale as the Zeeman energy at the applied magnetic field.
Therefore, it can be expected that the QCP of the AFM order appears at around 100 T and make the CO melting slow due to the critical slowing down.
Although the QCP in manganites have never been reported, we think that the effect of QCP becomes prominent in this study because of ultrahigh magnetic field.
In the field descending process, the critical field is far from the QCP of AFM order.
Therefore, the metamagnetic transition can occur sharply.

\begin{figure}
\centering
\includegraphics[width=0.95\columnwidth]{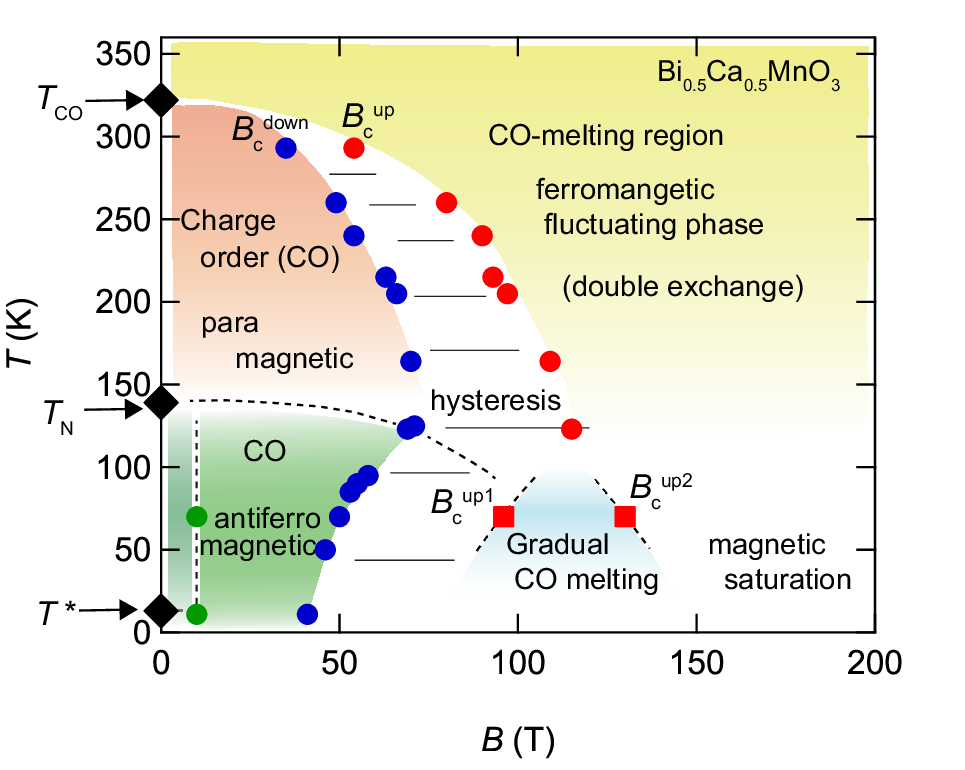}
\caption{\label{PhaDia} (Color online) $B$-$T$ phase diagram constructed. Electronic and magnetic states in the low-field region have been reported in several previous studies \cite{Bokov1967,RN99,RN100,RN77,RN105}.}
\end{figure}

Finally, we constructed the $B$-$T$ phase diagram of Bi$_{0.5}$Ca$_{0.5}$MnO$_3$, including the unusual critical feature at an ultrahigh magnetic field, as shown in Fig. \ref{PhaDia}.
The broken lines are the presumed phase boundaries.

For further investigation, we will measure other physical properties, such as X-ray diffraction, electrical resistivity, magnetostriction, and magneto-optical measurements under ultrahigh magnetic fields.

\section{Conclusion}
We investigated the magnetic properties of Bi$_{0.5}$Ca$_{0.5}$MnO$_3$ over a wide temperature range by means of magnetization measurements in ultrahigh magnetic fields up to 140 T generated by HSTC.
The metamagnetic transition with hysteresis was observed at room temperature, which indicates that the first order phase transition of field-induced CO melting occurs.
Meanwhile, the phase transition in the field ascending process cannot be detected below around $T_{\rm N}$,  whereas that in the field descending process appears.
This is probably understood as the gradual CO melting and its reformation.
The quenched spatial inhomogeneity of the charge distribution and/or AFM quantum criticality are possible causes to induce the gradual CO melting by a magnetic field.
Finally, we constructed the $B$-$T$ phase diagram, including the unusual critical features in ultrahigh magnetic fields.

\begin{acknowledgment}


We thank T. Nomura for experimental support and useful discussions, S. Imajo and K. Mitsumoto for fruitful discussions, T. Yajima for the PXRD experiment, and D. Hamane for the SEM experiments.

\end{acknowledgment}





\end{document}